\documentclass[aip,apl,twocolumn,reprint,superscriptaddress,showpacs,showkeys]{revtex4-1}

\usepackage{graphicx}
\usepackage{dcolumn}
\usepackage{bm}
\usepackage[utf8]{inputenc}
\usepackage[T1]{fontenc}
\usepackage{mathptmx}
\usepackage{amsmath}
\usepackage{color}
\usepackage{amssymb}

\begin{document}

\title{Overlap junctions for superconducting quantum electronics and amplifiers}

\author{Mustafa Bal}
 \email{mustafa.bal@nist.gov.}
 \affiliation{National Institute of Standards and Technology, Boulder, Colorado 80305, USA}
 \affiliation{Department of Physics, University of Colorado, Boulder, Colorado 80309, USA}
 
\author{Junling Long}
 \affiliation{National Institute of Standards and Technology, Boulder, Colorado 80305, USA}
 \affiliation{Department of Physics, University of Colorado, Boulder, Colorado 80309, USA}

\author{Ruichen Zhao}
 \affiliation{National Institute of Standards and Technology, Boulder, Colorado 80305, USA}
 \affiliation{Department of Physics, University of Colorado, Boulder, Colorado 80309, USA}
 
 \author{Haozhi Wang}
 \affiliation{National Institute of Standards and Technology, Boulder, Colorado 80305, USA}
 \affiliation{Department of Physics, University of Colorado, Boulder, Colorado 80309, USA}
 
 \author{Sungoh Park}
 \affiliation{National Institute of Standards and Technology, Boulder, Colorado 80305, USA}
 \affiliation{Department of Physics, University of Colorado, Boulder, Colorado 80309, USA}
 
  \author{Corey Rae Harrington McRae}
 \affiliation{National Institute of Standards and Technology, Boulder, Colorado 80305, USA}
 \affiliation{Department of Physics, University of Colorado, Boulder, Colorado 80309, USA}
 
  \author{Tongyu Zhao}
 \affiliation{National Institute of Standards and Technology, Boulder, Colorado 80305, USA}
 \affiliation{Department of Physics, University of Colorado, Boulder, Colorado 80309, USA}

 \author{Russell E. Lake}
 \affiliation{National Institute of Standards and Technology, Boulder, Colorado 80305, USA}
 \affiliation{Department of Physics, University of Colorado, Boulder, Colorado 80309, USA}

   \author{Daniil Frolov}
 \affiliation{Fermi National Accelerator Laboratory, Batavia, Illinois 60510, USA}
 
    \author{Roman Pilipenko}
 \affiliation{Fermi National Accelerator Laboratory, Batavia, Illinois 60510, USA}
 
   \author{Silvia Zorzetti}
 \affiliation{Fermi National Accelerator Laboratory, Batavia, Illinois 60510, USA}
 
   \author{Alexander Romanenko}
 \affiliation{Fermi National Accelerator Laboratory, Batavia, Illinois 60510, USA}
 
 \author{David P. Pappas}
 \email{David.Pappas@nist.gov.}
 \affiliation{National Institute of Standards and Technology, Boulder, Colorado 80305, USA}

\date{\today}

\begin{abstract}

Due to their unique properties as lossless, nonlinear circuit elements, Josephson junctions lie at the heart of superconducting quantum information processing. Previously, we demonstrated a two-layer, submicrometer-scale overlap junction fabrication process suitable for qubits with long coherence times. Here, we extend the overlap junction fabrication process to micrometer-scale junctions. This allows us to fabricate other superconducting quantum devices. For example, we demonstrate an overlap-junction-based Josephson parametric amplifier that uses only 2 layers. This efficient fabrication process yields frequency-tunable devices with negligible insertion loss and a gain of $\sim 30$ dB. Compared to other processes, the overlap junction allows for fabrication with minimal infrastructure, high yield, and state-of-the-art device performance.

\end{abstract}

\maketitle

Superconducting electronics have experienced rapid growth over the last decade. Advances in quantum information processing based on superconducting qubits has fueled much of this growth. Improvements in design, materials, development of quantum-limited amplifiers, and the implementation of low-loss superconducting components are some of the pillars supporting progress in the field.

Fundamental building blocks of superconducting quantum circuitry include discrete components, e.g., capacitors, inductors and Josephson junctions (JJs), as well as distributed element resonators. Modern microfabrication technology allows for the realization of these circuit elements with infrastructure that typically includes optical lithography techniques and metal/dielectric deposition. The JJs, formed by two superconducting electrodes separated by a thin tunnel barrier, play a central role in superconducting quantum circuits. The lossless, nonlinear inductance is the most salient property of these JJs, which render JJs as indispensable circuit elements for superconducting qubits as well as quantum-limited amplifiers.

There are various methods to realize JJs. These include subsequent layer removal of deposited superconductor-insulator-superconductor (SIS) tri-layers (i.e., the "trilayer process"),\cite{gurvitch1983,Weides2011} shadow evaporation (e.g., the Dolan-Bridges and Manhattan geometries),\cite{dolan1977,potts2001} and the overlap-junction technique.\cite{steffen2006,braumuller2015,wu2017} The trilayer process, typically used for superconducting electronics, involves many steps and significant infrastructure for different etches and both metal and dielectric deposition/patterning. On the other hand, shadow evaporation has become the norm for qubit JJs because they can be made very small with a single electron beam (e-beam) lithography step. This minimizes infrastructure and has enabled generations of researchers to make nano-scale quantum devices at the cost of scalability. However, with modern optical lithography that extends down to nanometer dimensions, this has become a moot point. Overlap junctions now allow access to full range of simple fabrication from micro- to nano-scale dimensions. This opens efficient wafer-level fabrication of JJs with high dimensional control and a wide range of applications.

Here, we develop low-critical-current JJs and demonstrate a simple process for micrometer-scale junctions. We describe the process bias and measure the current-voltage (I-V) characteristics of the resulting junctions. This enables the realization of a wide range of superconducting devices. In particular, we demonstrate the fabrication and characterization of important devices for the field, Josephson parametric amplifiers (JPAs), that use only two steps of photolithography. 

\begin{figure*}[ht!]
\includegraphics[width = 10.5 cm]{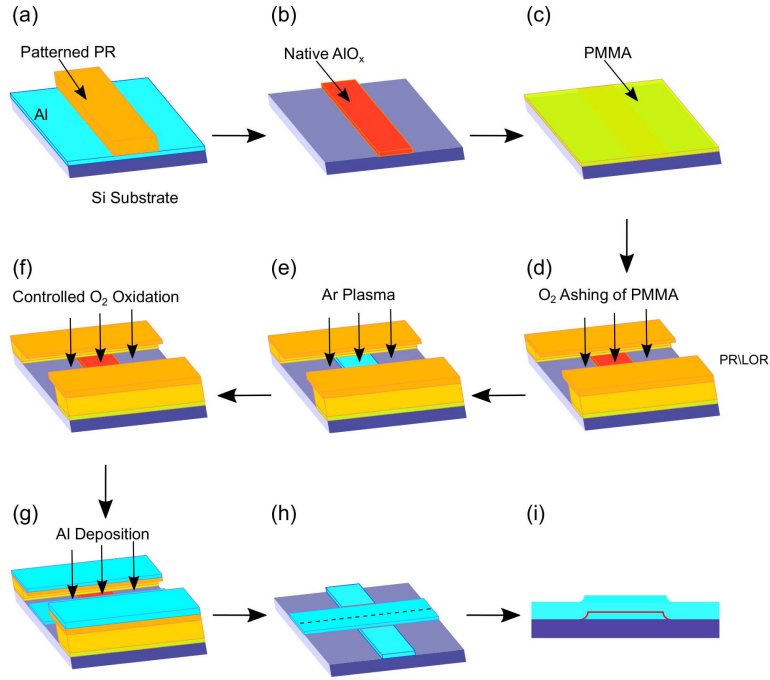}
\centering
\caption{Fabrication steps for micrometer-scale overlap junction process. 
	(a) 3D sketch of the Si substrate after deposition of the bottom Al layer followed by patterning the PR layer. 
	(b) After etching of Al bottom layer with etchant and stripping of the PR layer. A native surface oxide forms on the bottom Al electrode.
	(c) After coating with thin protective PMMA layer.
	(d) A bi-layer of PR/LOR is patterned in a second lithography step to define the pattern of the TE. The thin protective PMMA layer is removed by O$_{2}$ ashing at room temperature. 
	(e) The native oxide on the bottom Al electrode surface is removed by Ar RF-plasma cleaning.
	(f) Ultrahigh-purity grade low pressure O$_{2}$ is used to form a controlled AlO$_{x}$ tunnel barrier.
	(g) The TE is formed by subsequent Al evaporation. 
	(h) Metal-liftoff in solvent completes the overlap junction process.
	(i) A cross-section view of the junction along the dashed line in panel (h) of this figure.}
\label{Fig1}
\end{figure*}

As shown in Fig.~\ref{Fig1}, the process is started by depositing $\sim$~200~nm Al onto a 76 mm (3 in) intrinsic Si wafer with native oxide in an e-beam evaporation chamber at a base pressure of $\sim$~2.7 $\times$ 10 $^{-5}$~Pa. In the first step of photolithography, the desired pattern for the bottom-electrode (BE) layer is formed using a positive photoresist (PR) on the Al deposited wafer. The pattern is exposed through a photomask, i.e., reticle, in a 5:1 reduction stepper and developed in a tetramethylammonium hydroxide (TMAH) based solution (Fig.~\ref{Fig1}(a)). The pattern is then transferred onto the Al film by agitating the wafer in an etching solution heated to $\sim$~50~$^{\circ}$C (Fig.~\ref{Fig1}(b)). An over-etch of $\sim$~10~s ensures complete removal of Al in developed regions (unprotected by the PR).

In the second photolithography step, a metal lift-off resist (LOR) process was used (Fig.~\ref{Fig1}(c--i)). An imaging PR layer on top of the LOR layer comprises the bi-layer PR stacking. Since TMAH etches Al, a $\sim$~200~nm thick polymethyl methacrylate (PMMA) protective layer was applied, before the bi-layer PR stacking, to protect the BE Al layer from the PR developer. This process yields an undercut profile of $\sim$~0.7~$\mu$m after photo-exposure and development, resulting in a clean metal lift-off.

The subsequent steps shown in Fig.~\ref{Fig1}(d--f) are critical for the precise definition of the tunnel barrier area. Before the wafer is loaded into the high vacuum e-beam evaporator, it is subjected to room-temperature O$_{2}$ plasma ash (100~W for 3~min) at a pressure of $\sim$~67~Pa (Fig.~\ref{Fig1}(d)). This step removes PMMA and other organic residues in the developed regions of the wafer. Next, the wafer is loaded into the high vacuum deposition chamber and argon RF-plasma cleaning is employed to remove the native oxide on the patterned bottom Al layer as described in an earlier work.\cite{wu2017} This cleaning is carried out by applying RF power to an electrically isolated wafer holder that is water cooled. Typically 50~W RF power is applied for $\sim$1~min at room-temperature; the argon pressure is maintained at $\sim$~1.3~Pa (Fig.~\ref{Fig1}(e)). AlO$_{x}$ tunnel barrier is formed by controlled oxidation of the cleaned Al surface. (Fig.~\ref{Fig1}(f)). The oxidation is carried out by pumping out and sealing the chamber, then introducing ultra-high purity grade (99.999$\%$) O$_{2}$ into the chamber. The oxidation dose is tuned by adjusting the oxidation time and the oxidation pressure.\cite{zheng2015,Kleinsasser1995} The overlap junction is completed by depositing  $\sim 350$~nm Al as the top-electrode (TE) using e-beam deposition (Fig.~\ref{Fig1}(g)). The TE is created by lift-off in solvent with ultrasonic agitation (Fig.~\ref{Fig1}(h-i)).

An important step in fabricating tunnel junctions is the oxidation step (Fig.~\ref{Fig1}(f)). In our studies, we explored three different oxidation doses to vary the Josephson junction tunnel resistance (hence critical current) by nearly two orders of magnitude. The oxidation doses explored were 4~min at $\sim$~40~Pa, 36~min at $\sim$~40~Pa, and 36~min at $\sim$~100~Pa. 

\begin{figure}[t]
    \centering
    \includegraphics[width = 6.0 cm]{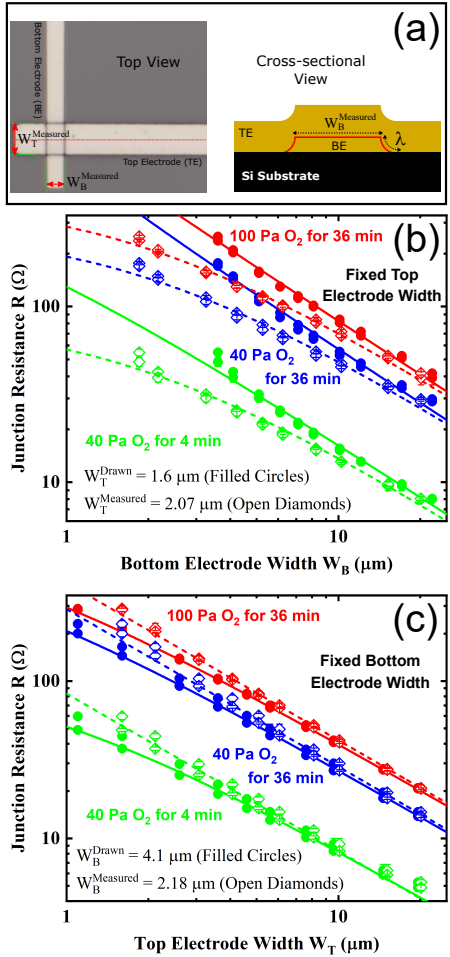}
    \caption{Room temperature junction characterization. (a) Optical microscope image of a typical micrometer-scale overlap junction is shown on the left of the panel. A drawing of the cross-section is presented on the right side. (b) Junction resistance $R$ vs $W_B$ at different oxidation doses as indicated on the plot. (c) Junction Resistance $R$ vs $W_T$.}
    \label{Fig2}
\end{figure}

\begin{table*}[t]
\begin{tabular}{||c|c||c|c|c||c|c||}
\hline
\multicolumn{2}{||c||}{Oxidation Dose} & \multicolumn{3}{c||}{Drawn Dimensions Model} &
\multicolumn{2}{c||}{Measured Dimensions Model} \\
\hline
$P$~(Pa)         & $t$~(min)        &$c~(\Omega~\mu$m$^{2}$)              & $a~(\mu$m)               & $b~(\mu$m)              &  $c~(\Omega~\mu$m$^{2}$)                         & $\lambda~(\mu$m)               \\
\hline
100 & 36 & 1634~$\pm$~45 & 0.40~$\pm$~0.03 & -0.11~$\pm$~0.09 & 1727~$\pm$~42 & 0.96~$\pm$~0.05\\
40 & 36 & 1087~$\pm$~55 & 0.34~$\pm$~0.06 & -0.19~$\pm$~0.17 & 1198~$\pm$~38 & 1.01~$\pm$~0.07\\
40 & 4  & 382~$\pm$~20  & 0.70~$\pm$~0.08 & 0.28~$\pm$~0.19  & 333~$\pm$~9   & 0.92~$\pm$~0.07\\
\hline
\end{tabular}
\caption{Oxidation dose and fitting results for models based on measured and drawn dimensions}
\label{table1}
\end{table*}

In order to design wafer-scale lithography patterns, it is necessary to understand the dimensional process biases and apply those to the drawn dimensions. In addition, for devices that are oxidized in-situ after argon cleaning, it is also necessary to include intrinsic process bias. These may be due to, e.g., inhomogeneous junction cleaning, oxidation, and extended edge profiles. To accomplish this, we plot and fit the room-temperature junction resistance $R$ as a function of both the drawn and the measured dimensions of the junctions. Making use of optical microscopy, we determined that the fabrication process yields actual dimensional biases of $a=+0.47\pm 0.03\  \mu$m and $b=-1.92\pm0.05\ \mu$m for the TE and BE, respectively. These reflect the fact that the top layer is additive and the bottom is subtractive. For the $R$ vs. electrode width fits, we start from $R \times A = c$ (where $A$ is the junction area and $c$ is a constant dependent on oxidation dose). To fit using the drawn areas, we can write:

\begin{equation}
R = \frac{c}{(W_{T}^{Drawn}+a)\times(W_{B}^{Drawn}+b)}
\label{eq1}
\end{equation}

\noindent Here $a$ and $b$ correspond to only dimensional bias, and $W_{T}^{Drawn}$ ($W_{B}^{Drawn}$) is the drawn width of top (bottom) electrode. The dimensional bias may be expected to correspond with the optically measured values if there is no intrinsic bias. However, as shown in Table~\ref{table1} and discussed below, we see that the fit bias for the BE is significantly smaller than the measured value, while they match fairly well for the TE. A fit using the measured dimensions helps to understand this; since the actual widths of the TE and the BE are known, we can fit it using only two parameters,

\begin{equation}
R = \frac{c}{W_{T}^{Measured}\times(W_{B}^{Measured}+2\lambda)}
\label{eq2}
\end{equation}

\noindent where again c is the oxidation constant while the parameter $\lambda$ includes intrinsic effects.

Resistance data was acquired for the three different oxidation doses and plotted vs. electrode widths, where first the $W_{B}$ was varied with fixed $W_{T}$ and then the $W_{T}$ is varied with fixed $W_{B}$ (Fig.~\ref{Fig2}). For Fig.~\ref{Fig2}(b) and (c), data for drawn vs. measured widths are plotted using solid and open symbols, respectively. Each data point in Fig.~\ref{Fig2}(b--c), represents the statistical average of 10 individual resistance measurements of nominally identical junctions across a die.  Measurements from two dies across the wafer are included for a given oxidation dose. For the various oxidation doses, we performed fits where $W_{B}$ is varied in Fig.~\ref{Fig2}(b) and $W_{T}$ is varied in Fig.~\ref{Fig2}(c). The dashed and solid lines in Fig.~\ref{Fig2}(b--c) show the fits using models based measured (Eq.~\ref{eq2}) and drawn (Eq.~\ref{eq1}) electrode dimensions, respectively.  The fitting results are summarized in Table~\ref{table1}.

For the resistance vs. measured width fits, we see that the intrinsic process bias on the bottom electrode, $2\lambda\sim 2\ \mu$m, accounts for the missing process bias in the parameter $b$ in the fit for the drawn dimensions. This could explicitly be accounted for in Eq.~\ref{eq1} by using:

\begin{equation}
b \rightarrow b+2\lambda
\label{eq3}
\end{equation}
 
Note that $\lambda$ is significantly larger than the BE thickness. Two possible explanations for this are (a) the edge profile is curved due to wet etching, which yields a length that is longer than the BE thickness and (b) RF plasma cleaning is localized towards the film edges, hence the edges are more electrically transparent than the top surface. This is by no means detrimental to the fabrication process and is easily accounted for in process bias or by extending the RF clean cycle time.\cite{RFclean}

The fitting parameter $c$ is a measure of normal resistance of the junction and is therefore proportional to the effective oxidation dose, given as $t^{\alpha}~\times~P^{\beta}$, where $t$ and $P$ are oxidation time and pressure, respectively. From the results presented in Table~\ref{table1}, $\alpha=0.58$ and $\beta=0.40$ ($\alpha=0.48$ and $\beta=0.44$) based on measured (drawn) dimensions. This result is in good agreement with accepted values in recent studies,\cite{zheng2015} illustrating that the oxidation process is not significantly altered by the air exposure and subsequent RF plasma cleaning step.

We next discuss the low temperature (40~mK) characteristics of the micrometer-scale overlap JJs. The current-voltage (I-V) characteristics are measured by driving current from a low-noise current source and measuring the voltage across the junction with a voltage pre-amplifier. 

\begin{figure}[ht]
    \centering
    \includegraphics[width = 6.0 cm]{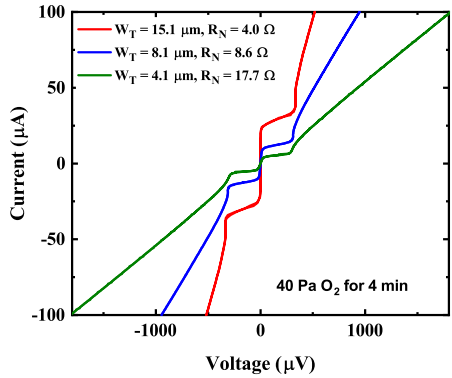}
    \caption{Low temperature characterization of single Josephson junctions. $W_{B}=2.18~\mu$m. $W_{T}$ and normal state resistance $R_{N}$ are shown on the plot. The measured superconducting gap is $2\Delta/e=335~\mu$V.}
    \label{Fig3}
\end{figure}

The JJs display I-V characteristics with no discernible hysteresis (Fig.~\ref{Fig3}). We acquire I-V curves of three representative junctions with $R_{N}$ values of $4.0, 8.6$ and $17.7\ \Omega$, with $W_{B}^{measured}=2.18~\mu$m. We calculate a superconducting gap  $2\Delta/e=335~\mu$V, in agreement with values expected for Al.\cite{Kittel1976} Critical currents, $I_{c}$, for each junction were 22, 10, and 4 $\mu$A. These measured $I_{c}$'s are significantly lower than expected, by about a factor of $1/3$, from the Ambegaokar-Baratoff (A-B) formula at $T=0$ K.\cite{AB1963} This is not surprising because the A-B formula is derived for an ideal uniform tunnel barrier, and  sets an upper limit for the maximum critical current. However, for an actual JJ, defects due to interface roughness, crystal and grain structures, impurities, and processing conditions all may lead to suppression of the critical current.\cite{Imamura1992,Shiota1992}

As a concrete and useful application of these JJs, we designed and fabricated JPAs around the overlap process. This process offers a method for making overlap JPAs (O-JPAs) reliably and with far fewer resources than are typically employed. Early research on JPAs dates back several decades.\cite{Calander1981,smith1985,yurke1989} More recently, JPAs are employed as indispensable tools for quantum information processing due to their ability to amplify small microwave signals with ultralow added noise. We refer to literature for various implementations and design aspects of JPAs.\cite{Manuel2008,Malnou2019,Hatridge2011,HSKu2014}

\begin{figure}
    \centering
    \includegraphics[width = 6.0 cm]{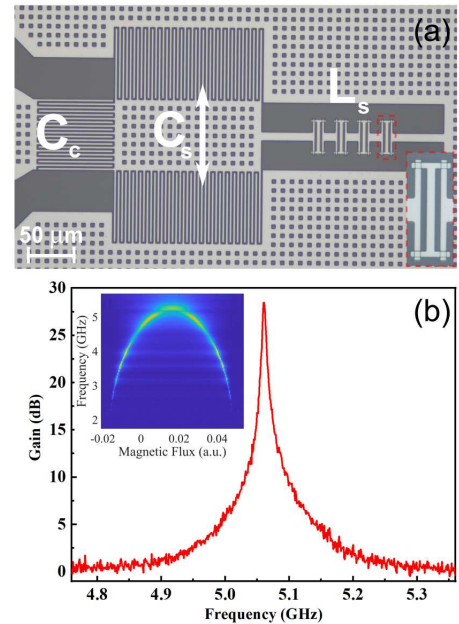}
    \caption{Physical layout and characterization of overlap junction based JPA. (a) The O-JPA is configured to be a lumped element $LC$ resonator using an interdigitated capacitor ($C_s=396$~fF) and a series array of 8~nominally identical SQUIDs with minimum $L_s^{min}=2.05$~nH as a tunable inductor. Lower right inset show the SQUID with $2.1 \times 2.2~\mu$m$^{2}$ overlap junctions. (b) Gain vs Frequency of O-JPA driven by a RF pump at 5.06 GHz. Inset shows frequency tunability as function of external magnetic flux.}
    \label{Fig4}
\end{figure}

Here, we implement an O-JPA design, which is comprised of a capacitive element in parallel with a non-linear inductor (Fig. \ref{Fig4}(a)).\cite{Malnou2019} The inductance is provided using a tunable superconducting quantum interference device (SQUID) array with inductance $L_s$ that is grounded on one end. For this device,  8~nominally identical SQUIDs, with junction inductance $L_j\approx0.5$~nH, provide a minimum $L_s^{min}=2.05$~nH  (inset of Fig. \ref{Fig4}(a)). An interdigitated shunt-capacitor, $C_s=396$~fF, is connected in parallel with the SQUID array to ground. The resulting lumped-element resonator is coupled to a coplanar waveguide (CPW) transmission line with a coupling capacitor $C_{c}=90$~fF.  The capacitors were made by etching a 2~$\mu$m wide trench between metal fingers.

The O-JPA is designed to have an unbiased resonance frequency of $f_{res}\approx5.05$~GHz with a quality factor of $Q_{Total}\approx32$.\cite{Sonnet} It is operated as a single port device in reflection mode. A circulator is used to separate the input signals (strong RF pump and weak RF signal) from the reflected output signals(idlers and amplified RF signal).

Initial tests of the O-JPA in an adiabatic demagnetization refrigerator ($T<100$~mK) are shown in Fig.~\ref{Fig4}, with the resonant frequency vs. SQUID bias in the insert of panel (b). The inductance of the SQUID array is tuned by applying an external magnetic flux to the SQUID loops by running a DC current through a coil placed nearby. This provides frequency tunability $\sim$ 3-5 GHz. A gain of more than 25 dB with about 7 MHz bandwidth is centered at 5.06 GHz. It is possible to engineer a wider bandwidth at the expense of lower gain. 

Cavity decay measurements at the base temperature of a dilution fridge (23 mK) allowed the initial estimate of the noise in O-JPAs.\cite{Romanenko2020} The decay of an ultra-high finesse cavity at 5 GHz was measured using an O-JPA, operated at 17 dB gain. The O-JPA was followed by a high-electron-mobility-transistor (HEMT) amplifier with 2.4 K thermal noise. The signal-to-noise ratio improved by  11 dB when the O-JPA is turned on, allowing the placement of a lower bound of half a photon added noise due to the O-JPA.

In conclusion, we developed a simple two-step process to realize micrometer-scale overlap JJ devices. The process requires minimal infrastructure compared to typical processes. The geometry, process, and oxidation dose are outlined and we show how to determine the process bias. We characterized the I-V curves of the JJs. We used this information to design and fabricate an ultra-low noise O-JPA with over 25 dB of gain.

\begin{acknowledgments}
We acknowledge support from the NIST Quantum Measurements Initiative and the U.S. Department of Energy through the Fermi National Accelerator Laboratory. R. Lake was supported by the NIST NRC Research Postdoctoral Associateship. We are very grateful for helpful discussions with  H.-S. Ku, X. Wu, G. Hilton, K. Lehnert, M. Malnou, and D. Palken during the initial stages of JPA design.
\end{acknowledgments}

The data that support the findings of this study are available from the corresponding author
upon reasonable request.

\nocite{*}
\bibliography{OverlapJPA}

\begin{thebibliography}{23}
\bibitem [{RFc()}]{RFclean}%
  \BibitemOpen
  \href@noop {} {}\bibinfo {note} {Argon RF-plasma cleaning at 50 W and 10
  mTorr are applied for 75 seconds for these devices. Since then we increased
  the RF cleaning time to at least 3 minutes to remove the oxide on top of the
  BE completely.}\BibitemShut {Stop}%
\bibitem [{Son()}]{Sonnet}%
  \BibitemOpen
  \href@noop {} {}\bibinfo {note} {We used high-frequency electromagnetic
  software for simulations.}\BibitemShut {Stop}%
\end{thebibliography}

\end{document}